\documentclass[12pt]{article}
\usepackage{graphicx}
\usepackage{amsmath}
\usepackage{amssymb}
\usepackage{url}
\newcommand\pubnumber{DPF2015-230}
\newcommand\pubdate{\today}

\def\msu{Department of Physics and Astronomy\\
Michigan State University, BPS Building, East Lansing, MI, USA}

\def\Title#1{\begin{center} {\Large #1 } \end{center}}
\def\Author#1{\begin{center}{ \sc #1} \end{center}}
\def\Address#1{\begin{center}{ \it #1} \end{center}}
\def\andauth{\begin{center}{and} \end{center}}

\newcommand\pubblock{\rightline{\begin{tabular}{l} \pubnumber\\
         \pubdate  \end{tabular}}}
\newenvironment{Abstract}{\begin{quotation}  }{\end{quotation}}
\newenvironment{Presented}{\begin{quotation} \begin{center} 
             PRESENTED AT\end{center}\bigskip 
      \begin{center}\begin{large}}{\end{large}\end{center} \end{quotation}}

\def\beq{\begin{equation}}
\def\eeq#1{\label{#1}\end{equation}}
\def\eeqn{\end{equation}}

\def\beqa{\begin{eqnarray}}
\def\eeqa#1{\label{#1}\end{eqnarray}}
\def\eeqan{\end{eqnarray}}

\let\bar=\overbar

\def\Dslash{\not{\hbox{\kern-4pt $D$}}}
\def\dslash{\not{\hbox{\kern-2pt $\del$}}}

\def\msb{{\bar{\ssstyle M \kern -1pt S}}}

\begin{document}

\pagenumbering{gobble}

\begin{titlepage}
\pubblock

\vfill
\Title{Searching for Primordial Black Holes with TeV Gamma Ray Detectors}
\vfill
\Author{J. T. Linnemann, D. Stump, S. S. Marinelli, T. Yapici, K. Tollefson}
\Address{\msu}
\Author{T. N. Ukwatta}
\Address{Space and Remote Sensing (ISR-2), Los Alamos National Laboratory, Los Alamos, NM 87545, USA}
\andauth
\Author{J. H. MacGibbon}
\Address{Department of Physics, University of North Florida, Jacksonville, FL 32224, USA}

\vfill
\begin{Abstract}
Primordial black holes (PBHs) are gravitationally collapsed objects that may have been created by density fluctuations in
the early universe and could have arbitrarily small masses down to the Planck scale. Hawking showed that due to quantum
effects, a black hole has a temperature inversely proportional to its mass and will emit all energetically allowed species of
fundamental particles thermally. PBHs with initial masses of order 5x10$^{15}$ g should be expiring in the present epoch with
bursts of high-energy particles, including gamma radiation in the GeV-TeV energy range. The Milagro high-energy
observatory, which operated from 2000 to 2008, is sensitive to the high end of the PBH evaporation gamma-ray spectrum.
A search of five years of Milagro data set a local (parsec-scale) upper limit of 3.6x10$^4$ PBH bursts/year/pc3. We will also
report the sensitivity of the High-Altitude Water-Cherenkov (HAWC) observatory to PBH evaporation events. Finally, we
investigate the final few seconds of black hole evaporation using Standard Model physics and calculate energy dependent
PBH burst time profiles in GeV/TeV range. We calculate PBH burst light curves observable by HAWC and explore search
methods and potentially unique observational signatures of PBH bursts.
\end{Abstract}
\vfill
\begin{Presented}
DPF 2015\\
The Meeting of the American Physical Society\\
Division of Particles and Fields\\
Ann Arbor, Michigan, August 4--8, 2015\\
\end{Presented}
\vfill
\end{titlepage}
\def\thefootnote{\fnsymbol{footnote}}
\setcounter{footnote}{0}

\pagenumbering{arabic}
\section{Introduction}

Primordial black holes (PBHs) are black holes created in the early
universe. Depending on the formation scenario, the PBHs could have masses ranging from the
Planck mass to more than million solar masses~\cite{Carr2010}. In 1974, Hawking showed by
convolving general relativity, thermodynamics and quantum field theory that a Black Hole
(BH) has a temperature inversely proportional to its mass and radiates with thermal spectra
photons and massive particles~\cite{Hawking1974}. As the BH emits this radiation,
its mass decreases and hence its temperature and emission rate increase.
A PBH that formed with an initial mass of $\sim 5.0 \times 10^{11}$ kg in the early
universe should be expiring today~\cite{MacGibbon2008} with a burst of high-energy particles,
including gamma-rays in the MeV to TeV energy range.
Thus PBHs are candidate gamma-ray burst (GRB) progenitors~\cite{Halzen1991}.

Confirmed detection of a PBH evaporation event would provide valuable insights into many areas
of physics including the early universe, high energy particle physics and the convolution
of gravitation with thermodynamics. Conversely, non-detection of PBH evaporation events
in sky searches would place important limits on models of the early universe.

The properties of the PBH final burst depend on the physics governing the production and
decay of high-energy particles. In the Standard Evaporation Model (SEM) which
incorporates the Standard Model of particle physics, a BH should directly Hawking-radiate
the fundamental Standard Model particles whose Compton wavelengths are of the order of the
black hole size~\cite{MacGibbon1990}. As the BH evaporates and loses mass over its lifetime,
its temperature surpasses the rest mass thresholds of further fundamental particle species.
When the temperature exceeds the Quantum Chromodynamics (QCD) confinement scale
($\sim 300\  {\rm MeV}$), the BH should directly emit quarks and gluons
which fragment and hadronize (analogous to jets seen in high-energy collisions in
terrestrial accelerators) as they stream away from the BH into the particles which are
stable on astrophysical timescales~\cite{MacGibbon1990, MacGibbon2008}. Thus according
to the SEM, the evaporating black hole will be seen astronomically as a burst of photons,
neutrinos, electrons, positrons, protons and anti-protons.

\section{Photons from a PBH Burst}\label{sec:pbh_theory}

\subsection{Hawking Radiation}\label{sec:Hawking_rad}

Hawking showed that a black hole radiates fundamental particle species at
an emission rate of~\cite{Hawking1974, Carr2010}
\begin{equation}\label{eq:hawking}
\frac{d^{2}N}{dE dt} = \frac{\Gamma/2\pi\hbar}{e^{x} - (-1)^{2s}} \ n_{\rm dof} ,
\end{equation}
where $s$ is the particle spin and $n_{\rm dof}$ is the number of degrees of freedom of the
particle species (e.g. spin, electric charge, color and flavor).
The dimensionless quantity $x$ is defined by
\begin{equation}\label{eq:xeq}
x \equiv \frac{8\pi G M E}{\hbar c^{3}} = \frac{E}{kT_{BH}}
\end{equation}
where $E$ is the energy of the emitted particle, $M$ is the black hole mass,
$T_{BH}\propto 1/M$ is the black hole temperature, and $G$, $c$, and $\hbar$
are the universal gravitational constant, speed of light, and the reduced
Planck constant, respectively. The absorption coefficient $\Gamma$ depends
on $M$, $E$ and $s$. For an emitted species of rest mass
$m$, $\Gamma$ at $E \gg mc^{2}$ has the form
\begin{equation}\label{eq:Page}
\Gamma(M, E, s) = 27 \left(\frac{x}{8\pi}\right)^{2} \gamma_{s}(x)
\end{equation}
such that $\gamma_{s}(x)\rightarrow 1$ for large $x$.

To calculate the spectrum of the final photon burst from the PBH, two important relations
pertaining to the final phase of BH evaporation are needed. The first relation is the black
hole mass $M$ expressed as a function of remaining lifetime. The mass loss rate can be written
as~\cite{Halzen1991}
\begin{equation}\label{eq:massloss}
\frac{dM}{dt}\equiv -\frac{\alpha(M)}{M^{2}},
\end{equation}
where the factor $\alpha(M)$ incorporates all emitted particle species and degrees of freedom.
As the BH evaporates, the value of $M$ is reduced by an amount equal to the total mass-energy
of the emitted particles. By conservation of energy,
\begin{equation}\label{disc:conserve_energy}
\frac{d(Mc^2)}{dt} = - \sum_i \int_0^\infty \frac{d^2 N_i}{dEdt} E dE
\end{equation}
where the summation over $i$ is over all radiated species. Therefore,
\begin{equation}\label{disc:alpha2}
\alpha(M) = \frac{M^2}{c^2} \sum_i \int_0^\infty \frac{d^2 N_i}{dEdt} E dE.
\end{equation}
Assuming the Standard Model including the top quark, 125 GeV Higgs, and 3 families of
massive Majorana neutrinos, we find that the asymptotic value $\alpha_{\rm SM}$
of $\alpha(M)$ as $M$ decreases is
\begin{equation}
\alpha_{\rm SM} = 8.40 \times 10^{17}\, \rm kg^{3} s^{-1}.
\end{equation}
For the current and future generations of high energy gamma-ray observatories,
we are interested in bursts generated by black holes of
temperature $T_{BH}\gtrsim 1\, {\rm TeV}$. Thus for $T_{BH} \gtrsim 1\, {\rm TeV}$ BHs
(corresponding to $M \lesssim 10^{7} \ {\rm kg}$ and a remaining evaporation
lifetime of $\tau \lesssim 500\,{\rm s}$), $\alpha(M) \approx \alpha_{\rm SM}$ and the
BH mass as a function of remaining evaporation lifetime $\tau$ is
\begin{equation}\label{eq:masstau}
M(\tau) \approx \left(3\alpha_{\rm SM}\, \tau\right)^{1/3} =
1.36 \times 10^{6}\  \left(\frac{\tau}{1 s}\right)^{1/3}\ {\rm kg}.
\end{equation}

The second relation that we require for the final evaporation phase is the temperature $T_{BH}$
expressed as a function of $\tau$. Combining Equations \ref{eq:xeq} and \ref{eq:masstau}, we have
\begin{equation}\label{eq:temptau}
kT_{BH} = \frac{\hbar c^{3}}{8\pi G M}
= 7770\ \left(\frac{1 s}{\tau}\right)^{1/3} \ {\rm GeV}.
\end{equation}
for $k T_{BH}\gtrsim 1\, {\rm TeV}$.

\subsection{QCD Fragmentation}

According to the SEM, Equation \ref{eq:hawking} describes the direct Hawking radiation of
the fundamental particle species of the Standard Model: the leptons, quarks, and the gauge
bosons~\cite{MacGibbon1990, Halzen1991}. As they stream away from the BH, these particles
will then evolve by Standard Model processes, ultimately into the particles which
are stable on astrophysical timescales: photons, neutrinos, electrons, positrons,
protons, and antiprotons. In particular, quarks and gluons will undergo fragmentation and hadronization
into intermediate states which will eventually decay into the astrophysically stable particles.

For application to PBH searches at gamma-ray observatories, we seek the total photon
emission rate from the BH. The photon spectrum has several components:
(i) the ``direct photons'' which are directly Hawking radiated by the BH:
this component peaks at a few times $T_{BH}$ and is most important at the highest
photon energies at any given $T_{BH}$; (ii) the ``fragmentation photons'' arising
from the fragmentation and hadronization of the directly Hawking radiated quarks and gluons:
this component is the dominant source of photons at energies below $T_{BH}$; and
(iii) the photons produced by the decays of other Hawking-radiated fundamental particles,
e.g. the tau lepton, $W$ and $Z$ gauge bosons, and Higgs bosons:
this component is small compared to the component produced by the directly Hawking radiated
quarks and gluons and is neglected here. (We note that, because the $W$, $Z$, and Higgs bosons
decay predominantly via hadronic channels, the main effect of component (iii) will be to enhance
the fragmentation photon component (ii) somewhat.)

\subsection{The Pion Fragmentation Model}

In jet fragmentation and hadronization, most of the photons arise from the decays of
$\pi^{0}$ states, which is $\pi^{0}$  decays to 2$\gamma$'s with a branching fraction 98.8\%.
We proceed assuming that the QCD fragmentation of quarks and
gluons may be approximated entirely by pion production. In our pion fragmentation model,
two questions must be addressed: what is the pion distribution generated by the partons
(the initial quarks and gluons) and what is the photon spectrum generated by the pion decays?

To answer the first question, we utilize a heuristic fragmentation function
\begin{equation}\label{eq:hff}
D_{\pi /i}(z)=\frac{15}{16}\ z^{-3/2}\ (1-z)^{2}
\end{equation}
where $z \equiv E_{\pi}/E$ is the energy fraction carried by a pion
generated by a parton of energy $E$~\cite{Halzen1991}.
We assume the same form of $D_{\pi /i}(z)$ for all partons and that all of the parton
energy goes into pions i.e. the function in Equation \ref{eq:hff} is normalized such that
$\int_{0}^{1} z D_{\pi /i}(z) dz = 1$. The instantaneous BH pion production rate per
pion energy interval is then
\begin{equation}\label{eq:d2Npi}
\frac{d^{2}N_{\pi}}{dE_{\pi} dt} =
\sum_{i} \int_{m_{\pi}c^{2}}^{\infty} \int_{0}^{1}
 \frac{d^{2}N_{i}}{dE dt}\ D_{\pi /i}(z)\ \delta(E_{\pi}-z E) dz dE.
\end{equation}

\subsection{Photon Flux from Pion Fragmentation}

To answer the second question, we obtain the photon flux from the $\pi^{0} \rightarrow 2\gamma$ decay of the
pion distribution. Because the fragmentation function $D_{\pi /i}(z)$ includes
all three pion charge states $\pi^{+}, \pi^{-},$ and $\pi^{0}$ as equal components,
and each $\pi^{0}$ decays into two photons, we must multiply by 2/3 to get the
$\gamma$ multiplicity. In the $\pi^{0}$ rest frame, the two photons have equal energies,
$m_{\pi}c^2/2$ and equal but opposite momenta. In the reference frame of the gamma-ray detector,
the energies of the two photons are unequal but complementary fractions
of the $\pi^{0}$ energy in the detector frame, $E_{\pi}$.
We assume that only one of the photons in each pair is detected.\footnote{
The angle between the 2 photon trajectories in the detector frame will be very small because
of the large Lorentz boost.
However, if the BH is at a distance of order 1 parsec from the detector,
then only one of the photons from each $\pi^0$ decay will hit the detector.}

Let $\theta^{\prime}$ be the angle between the momentum of the observed photon
in the $\pi^{0}$ rest frame and the momentum of the pion in the detector frame.
In the detector frame, the photon energy $E_{\gamma}=(E_{\pi} /2)(1+\beta\cos\theta^{\prime})$
where the $\pi^{0}$ velocity $\beta = v/c\ \approx 1$
and $E_{\pi} = m_{\pi}/\sqrt{1-\beta^2}$. Because the angular distribution of the photons is
isotropic in the $\pi^{0}$ rest frame,
the distribution of photon energy in the detector frame is
\begin{equation}\label{eq:frag}
\frac{d^{2}N_{\gamma}}{dE_{\gamma} dt} =
\frac{2}{3} \int_{-1}^{1} \frac{2\pi d\cos{\theta^{\prime}}}{4\pi}
\int_{m_{\pi}}^{\infty}
\frac{d^{2}N_{\pi}}{dE_{\pi} dt}
\delta[E_{\gamma}-(E_{\pi}/2)(1+\beta \cos\theta^{\prime})] dE_{\pi} .
\end{equation}

\subsection{Parametrization of the Direct and the Fragmentation Contribution}
To further simplify our analysis and subsequent calculations, we parameterize
the directly Hawking radiated and the pion fragmentation components of the
instantaneous PBH burst photon spectrum using the variable
\begin{equation}
\xi_\gamma = 1.287 \times 10^{-4} \left(\frac{E_\gamma}{1 \rm{GeV}}\right) \left(\frac{1 s}{\tau}\right)^{1/3}
\end{equation}
as follows:

\begin{equation}\label{eq:pbh_inst_para_1}
\left(\frac{d^{2}N_{\gamma}}{dE_{\gamma}\,dt}\right)_{\rm direct} = 1.13 \times 10^{19} (\xi_\gamma)^6 (\exp(\xi_\gamma) - 1)^{-1} \times F(\xi_\gamma) \ \rm{GeV}^{-1}\ \rm{s}^{-1}
\end{equation}
where
\begin{eqnarray}
F(\xi_\gamma)& = & 1.0 \ \ \text{for} \ \xi_\gamma \le 2\nonumber \\
F(\xi_\gamma)& = & \exp([-0.0962-1.982(\ln \xi_\gamma - 1.908)][1+\tanh[20(\ln \xi_\gamma - 1.908)]])\ \ \text{for} \ \xi_\gamma > 2\nonumber
\end{eqnarray}
and
\begin{eqnarray}\label{eq:pbh_inst_para_2}
\left(\frac{d^{2}N_{\gamma}}{dE_{\gamma}\,dt}\right)_{\rm frag.} & = & 6.339 \times 10^{23} \ (\xi_\gamma)^{-3/2} [1 - \Theta_{S}(\xi_\gamma-0.3)] \nonumber \\
 & + & 1.1367 \times 10^{24} \ \exp(-\xi_\gamma) [\xi_\gamma (\xi_\gamma + 1)]^{-1} \ \Theta_{S}(\xi_\gamma-0.3)\ \rm{GeV}^{-1}\ \rm{s}^{-1}
\end{eqnarray}
where
\begin{eqnarray}
\Theta_{S}(\lambda) & = & 0.5 [1 + \tanh(10 \ \lambda)]. \nonumber
\end{eqnarray}
These parametrizations are valid for energies above $E_{\gamma}\sim 1$ GeV and are accurate to $\pm$ 15 \%
for $\xi_\gamma$ in the range from 0.1 to 5.0, and to $\pm$ 3 \% for smaller and larger $\xi_\gamma$.
If greater accuracy is required, we use a table of the ratios of the exact integration of Equation~\ref{eq:d2Npi}
using the fragmentation function~\ref{eq:frag} to the parameterized value, to correct the parameterized values.

\subsection{PBH Burst Light Curve}\label{subsec:pbh_lightcurve}

To find the time evolution of the PBH burst, we integrate the total instantaneous photon
flux $d^{2}N_{\gamma}/(dE_{\gamma} dt)$ over photon energy $E_{\gamma}$
while retaining the time dependence:

\begin{equation}\label{eq:LightCurve}
\bigg[\frac{dN_{\gamma}}{dt}\bigg]_{\text{Emission}}=\int_{E_{\rm min}}^{E_{\rm max}} \frac{d^{2}N_{\gamma}}{dE_{\gamma} dt} dE_{\gamma}.
\end{equation}
\\
In general $E_{\rm min}$ and $E_{\rm max}$ are set by the energy range of the detector.
As a case study, we use here the High Altitude Water Cherenkov (HAWC) observatory~\cite{Ukwatta2015}
for which the relevant energy range is $E_{\rm min}$=50 GeV  to $E_{\rm max}$=100 TeV.

\begin{figure}
\begin{center}
\includegraphics[height=2.5in]{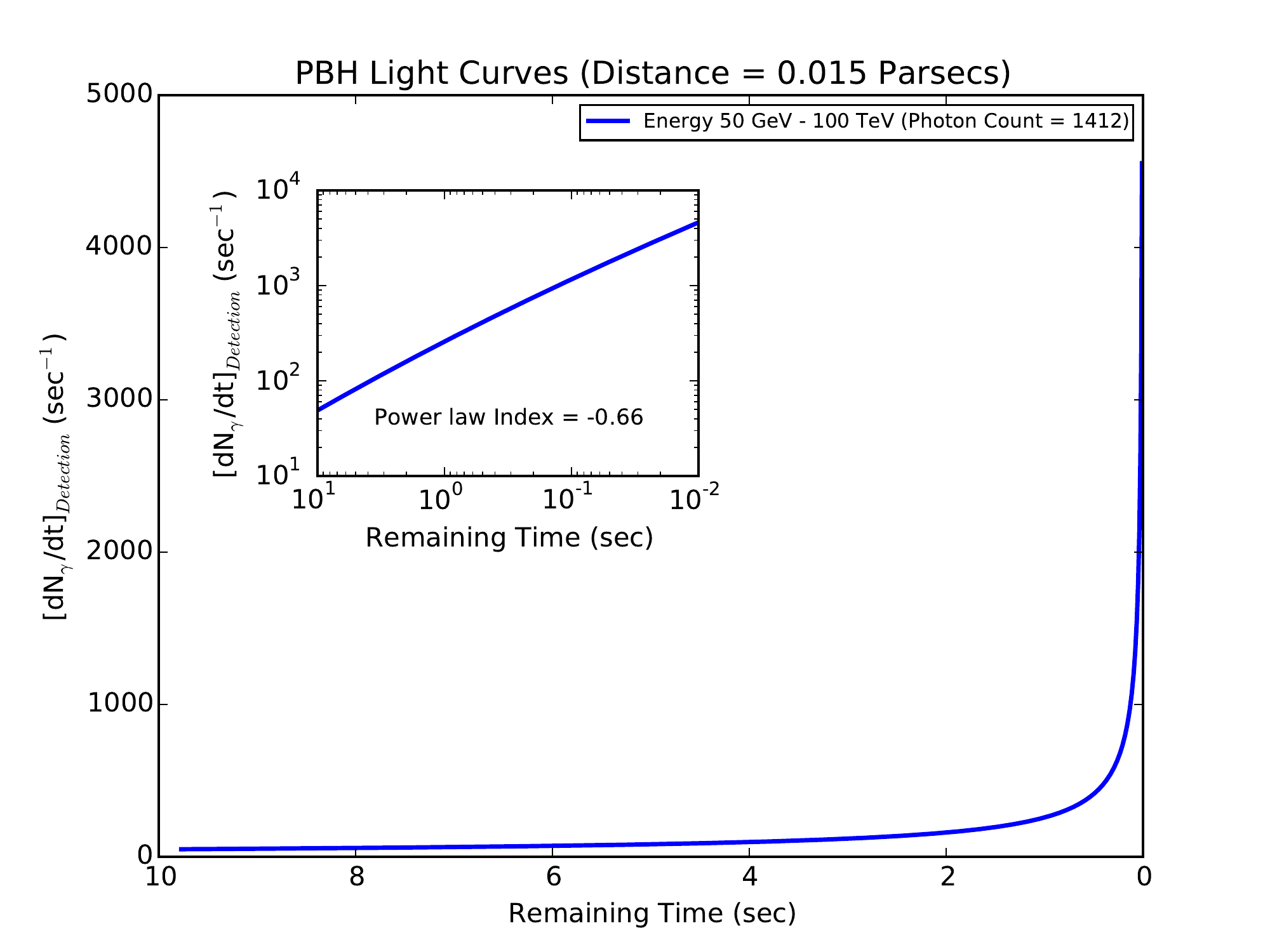}
\caption{Simulated PBH Burst lightcurve observed by HAWC (at a distance of 0.05 light years)
obtained by convolving with the HAWC effective area published in Ref~\cite{Ukwatta_ICRC_PBH_2013}.
This shape is also well described by a power law with a index of -0.66.
\label{fig:pbh_lightcurve_hawc1}}
\end{center}
\end{figure}

The time profile (the {\em light curve}) of the PBH burst at the detector can be calculated as follows:
\begin{equation}\label{eq:DetectorLightCurve}
\bigg[\frac{dN_{\gamma}}{dt}\bigg]_{\text{Detection}}=\frac{1}{4\pi r^{2}}
\int_{E_{\rm min}}^{E_{\rm max}} A(E_{\gamma})
\frac{d^{2}N_{\gamma}}{dE_{\gamma} dt} dE_{\gamma}
\end{equation}
where $A(E_{\gamma})$ is the effective area of the HAWC detector and $r$ is the distance
to the PBH. Figure~\ref{fig:pbh_lightcurve_hawc1} shows the time profile of a
PBH burst at a distance of 0.05 light years in the HAWC energy range and for
the HAWC effective area published in Ref~\cite{Ukwatta_ICRC_PBH_2013}.

\begin{figure}
\begin{center}
\includegraphics[height=2.5in]{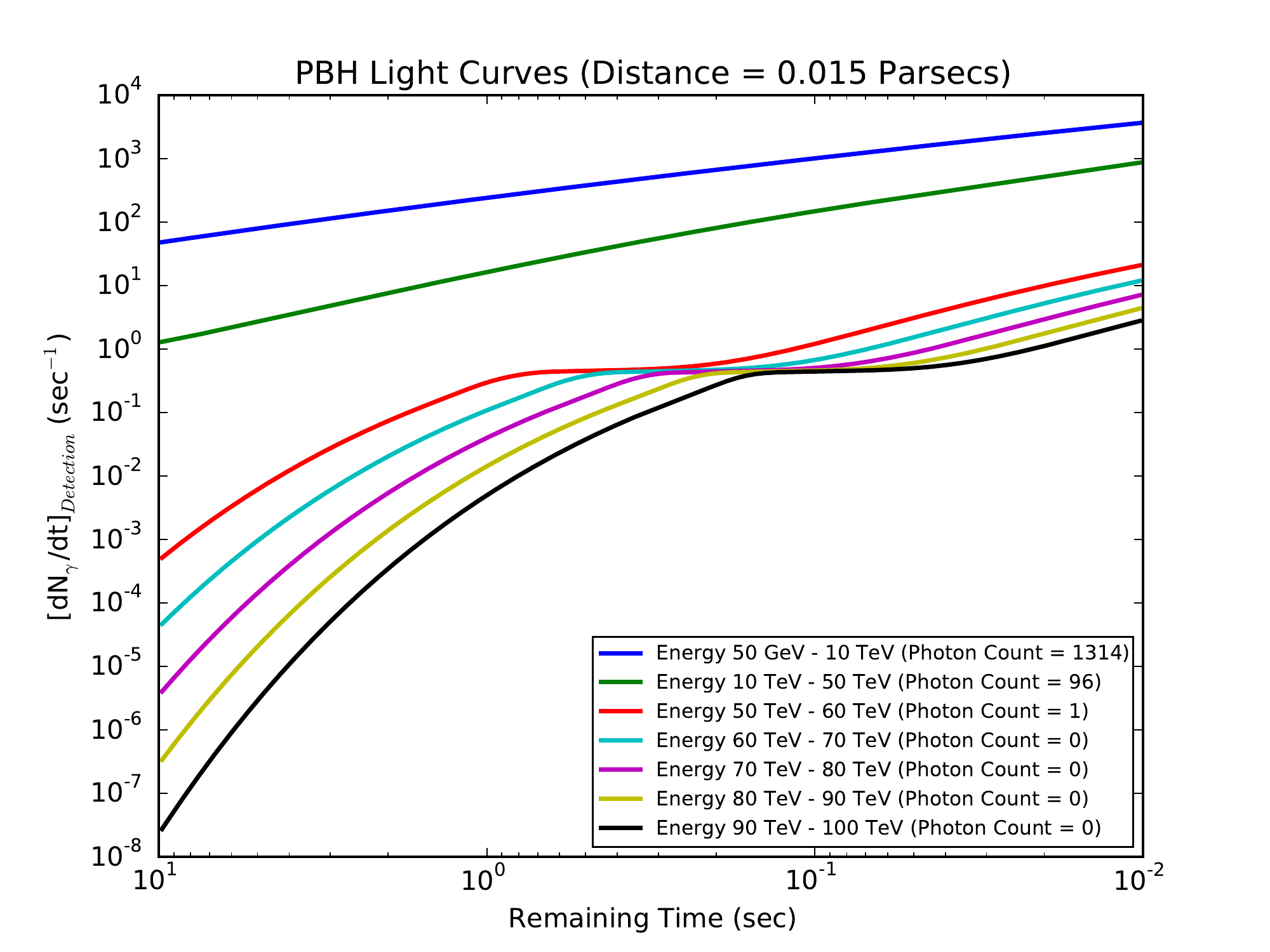}
\caption{Simulated detection time profiles of a PBH burst at a distance of 0.05 light years observed by HAWC in multiple energy bands.
Photon numbers detected in each energy band is shown in the legend.
\label{fig:pbh_lightcurve_hawc2}}
\end{center}
\end{figure}

It is also interesting to investigate the energy dependence of the PBH burst time profile.
In Figure~\ref{fig:pbh_lightcurve_hawc2}, we show $dN_{\gamma}/dt$ calculated using
various energy bands in the HAWC energy range. As seen in Figure~\ref{fig:pbh_lightcurve_hawc2},
the low energy bands show similar emission profiles. However, above $E_{\gamma}\sim$ 10 TeV
the burst emission time profile is energy-dependent and exhibits an inflection
around $\tau \sim 0.1$ seconds. This occurs due to the domination of the direct
photon component at the highest energies.

\section{Summary and Conclusions}\label{sec:conclusion}

In this paper, we have briefly reviewed the theoretical framework of the PBH standard
emission model and calculated the PBH burst light curves which would be observed by HAWC.
The main findings and conclusions of our paper are:

\begin{enumerate}
 \item We have developed approximate analytical formulae for the instantaneous
 PBH spectrum which includes both the directly Hawking radiated photons and the photons arising from the other directly Hawking radiated species.

 \item For the first time, we have calculated the PBH burst light curve and studied
 its energy dependence at a detector.

 \item For low energies ($E_{\gamma}< 10$ TeV) the light curve profile does not show much
 evolution with energy and is well described by a power law $\tau^{-0.66}$.
 However, at high energy the light curve displays significant energy dependence
 that may be used as an unique signature of PBH bursts. The HAWC observatory is sensitive in
 this high energy range and potentially can be used to uniquely identify PBH bursts.
\end{enumerate}

\section*{Acknowledgments}
\footnotesize{ This work was supported by grants from the National Science Foundation
(MSU) and Department of Energy (LANL). TNU acknowledges the support of this work by the
Laboratory Directed Research \& Development (LDRD) program at LANL. We would also like to thank
Alberto Carrami{\~n}ana of INAOE (National Institute of Astrophysics, Optics and Electronics
) in Mexico and Pat Harding at LANL for useful feedback on the draft of the paper.
}

\def\aj{\emph{AJ.}}
\def\APJ{\emph{ApJ.}}
\def\apj{\emph{ApJ.}}
\def\aap{\emph{A.\& A.}}
\def\apjs{\emph{ApJS}}
\def\apss{\emph{Ap\&SS}}
\def\apjl{\emph{ApJ. Lett.}}
\def\araa{\emph{ARA\&A}}
\def\mnras{\emph{MNRAS}}
\def\nat{\emph{Nature}}
\def\NAT{\emph{Nature}}
\def\pre{\emph{Phys. Rev. E}}
\def\prd{{\em Phys. Rev.} D}
\def\pasj{\emph{PASJ}}
\def\jcap{\emph{JCAP}}
\def\physrep{\emph{Phy. Rep.}}

\end{document}